\documentclass{article}
\usepackage{cite}
\usepackage{spconf,amsmath,graphicx}
\usepackage{amssymb,amsfonts}
\usepackage{textcomp}
\usepackage{xcolor}
\usepackage{booktabs}
\usepackage{multirow}
\usepackage{url}
\usepackage[table]{xcolor}
\usepackage{graphicx}
\usepackage{tikz}
\usepackage[skip=1pt plus1pt, indent=1em]{parskip}
\usetikzlibrary{positioning}

\definecolor{slate}{HTML}{2A3442}
\definecolor{coral}{HTML}{E2674A}
\definecolor{coralbg}{HTML}{FBE7E0}
\definecolor{bluesup}{HTML}{3E7CB1}
\definecolor{bluebg}{HTML}{E4EEF6}
\definecolor{teal}{HTML}{2A9D8F}
\definecolor{tealbg}{HTML}{E2F3EF}
\definecolor{labelgrey}{HTML}{5B6B7B}
\definecolor{cardborder}{HTML}{C9D2DC}
\definecolor{greyborder}{HTML}{B9C2CC}
\definecolor{greybg}{HTML}{ECEFF2}
\definecolor{greytext}{HTML}{4A5663}

\title{CARES: A Controlled Synthetic Benchmark of Speaker Reactions to Sound}
\name{Marcel Gibier$^{1,2}$, Thomas Thebaud$^{2}$, Olivier Boëffard$^{2}$, Jean-Fran\c{c}ois Bonastre$^{2}$}
\address{
  $^{1}$Inria, Paris, France\\
  $^{2}$AMIAD, France\\
  \texttt{marcel.gibier@polytechnique.edu}
}
\begin{document}
\ninept
\maketitle
\begin{abstract}
Automatic audio scene description turns a recording into a text account
of a situation. One difficulty is deciding which elements of the audio should be kept, since a description cannot include them all. Annotators disagree about it, making a ground truth hard to obtain. In this work,  we first define the ground truth, then generate the data. We focus on audio events and define sound salience with a simple rule: a sound is salient when a speaker audibly reacts to it. For scale and variety, a controlled set of scenarios fixes the ground
truth and a language model writes the dialogues. The resulting corpus, \textsc{CARES}, contains 10,000 two-speaker scenes. We then benchmark six audio–language models on three tasks: identifying the scene, tagging the
sounds present, classifying reactions.  We show that these models hear the sounds but miss how the speakers react to them.
\end{abstract}

\begin{keywords}
Audio scene description, rich transcription, speech summarization,
synthetic dataset, speaker reaction
\end{keywords}

\section{Introduction}
\label{sec:intro}

An audio scene is the audio trace of a real situation, bounded in time
and place, like a scene in a film. It reaches the microphone as one
mixed stream, where speech, music and environmental sounds share a
single timeline. Automatic audio scene description (AASD) turns this
stream into text: who is present, what they do and say, what happens
around them, and how they relate over time.

AASD raises three difficulties. First, the information is spread
across all the layers of the recording: who speaks, what is said and
how, the events, the ambience, the music. These layers only make sense
together. A ring tone alone means little, and so does a speaker
breaking off; jointly, they are an interrupted meal. Second, a
description cannot keep everything. Its length is a choice, from a full transcript to a few sentences, depending on what the description is for. Choosing it means choosing which
elements are salient, that is, worth keeping. This is not perceptual
salience, where a sound stands out from the
background~\cite{kaya2017}. The loudest sound is not always worth
reporting; a quiet phone buzz that stops a conversation is. Third, the
task is hard to evaluate: no single description is correct, and
annotators asked to select salient content agree only
weakly~\cite{martinmorato2021macs}.

\begin{figure}[t]
\centering
\includegraphics[width=0.98\linewidth]{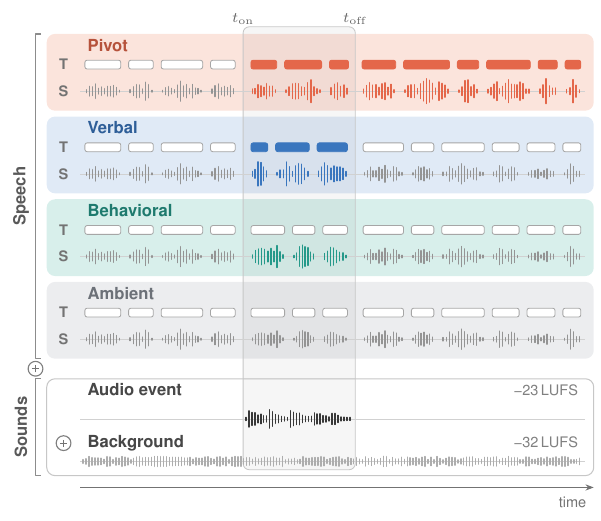}
\caption{Four types of speaker reaction to an audio event. Each block shows the transcript
(T, one box per word) and the speech signal (S) over time. The shaded box marks the
audio event $[t_{\mathrm{on}}, t_{\mathrm{off}}]$. Colored elements are altered by the
event. \textbf{Pivot}: persistent change of T and S. \textbf{Verbal}: local verbal
reaction (T and S). \textbf{Behavioral}: local change of S only. \textbf{Ambient}: no
impact. The speech signal, the audio event ($-23$\,LUFS) and the background
($-32$\,LUFS) are summed ($\oplus$) to form the final mix.}
\label{fig:taxonomy}
\end{figure}

Prior work sits at the two ends of this length scale. Rich
transcription enriches the words with speaker turns, sentence units,
disfluencies and non-speech events~\cite{gravier2004ester}. It weights
nothing, so salience is left to the user. Speech summarization keeps
only the parts of the transcript it scores as
important~\cite{furui2004speech}: the system decides. Both focus on
speech and treat environmental sounds as side information.

Recent audio--language models remove this
separation~\cite{goel2026afnext,xu2025qwen3omni,kimiteam2025kimiaudio,xiaomi2025mimoaudio,yang2026mossaudio,dinkel2025midasheng}.
They take the mixed stream as input and can, in principle, relate what
is said to what is heard. This makes evaluation critical: a model can
transcribe a scene correctly yet miss the sound the speakers reacted
to, or report sounds that were never
there~\cite{kuan2024understanding}. Current datasets record
what was said or heard, not which sounds mattered to the participants.

What is needed is a large set of scenes in which speech and sounds come
with an unambiguous ground truth. We therefore reverse the order of construction. We first define
salience and fix the ground truth from it. Only then are the scenes
generated artificially. We restrict salience to audio events, not what
is said, and define it through the talk: an audio event is salient when
a speaker audibly reacts to it (Fig.~\ref{fig:taxonomy}). This simple
definition can later be extended.

We address two questions. (Q1) Can we generate audio scenes whose salience
is known in advance? Yes: Our generated corpora, \textsc{CARES} (\textit{Controlled Audio
Reactions in Environmental Scenes}), contains 10{,}000 two-speaker
scenes associated with a defined sound salience. And both human annotators ($\kappa = 0.86$ on 200 scenes) and an independent model recover the salience information from the transcript.
(Q2) Do audio--language
models take this salience into account? Our benchmark shows they do
not: they recognize the sounds, but tell how speakers react no better
than a text-only system.

\section{Related Work}
\textbf{Describing an audio scene.} An audio scene has several components, including speech, environmental sounds, and music. Most tasks describe only one of them at a time. Audio captioning pairs short clips with a free-form sentence~\cite{kim2019audiocaps}. Sound event detection and acoustic scene classification label which sounds occur and when~\cite{gemmeke2017audioset}. Meeting corpora~\cite{carletta2005ami} and spoken dialogue summarization describe what was said\cite{lu2025spokendialogsum}. Synthetic conversations have been mixed with background sounds, but only to make
the audio realistic~\cite{labrak2026synthdopaco}.

\textbf{Labels by construction.} Synthetic soundscapes reverse the order
of annotation: Scaper~\cite{salamon2017scaper} mixes isolated events so
that their class and onset are known exactly.

\textbf{Sounds in spoken interaction.} Audio--language models take
speech and sounds in one input. Benchmarks question them on existing
recordings~\cite{sakshi2025mmau}, or test speech, scenes and
events as separate targets~\cite{sseubench2025}. Spoken dialogue
benchmarks add background sounds that the system's reply should take
into account, for single turns~\cite{ao2024sdeval}, multi-turn
dialogues~\cite{cheng2025voxdialogue}. Closest to us,
CASU~\cite{zhang2026casu} mixes synthetic speech (a monologue or a
two-person exchange, under 30\,s) with retrieved events and
backgrounds, and asks multiple-choice questions: whether the sounds
confirm, contradict or contextualize what is said, which entity or role
they reveal, and how an interpretation would change if they differed.
The sound provides evidence for the listener, such as when a speaker sounds confident while an event proves them wrong. Whether the speakers themselves notice or react to the sound is not labelled.

\textbf{Reactions to sounds.} Conversation analysis has studied both
of our axes through detailed analysis of recorded interactions, without
labelled data: how people notice events around them
~\cite{keisanen2012}, and how they cut off or adapt their talk
~\cite{schegloff1977}. Rich transcription marks
disfluencies and non-speech events~\cite{gravier2004ester}, but never
the link between them. CARES labels this link, at scale.

\section{Method}
\label{sec:method}
Our aim is to generate a large and varied set of two-speaker scenes in
which the salience of every audio event is known before the audio
exists. We use a language model only for the parts that require natural
language, and generate everything else deterministically
(Fig.~\ref{fig:pipeline}).

\subsection{Templates and scenarios}
\label{ssec:scenarios}
\begin{figure*}[t]
    \centering
    \makebox[\textwidth][c]{\includegraphics[width=0.95\textwidth]{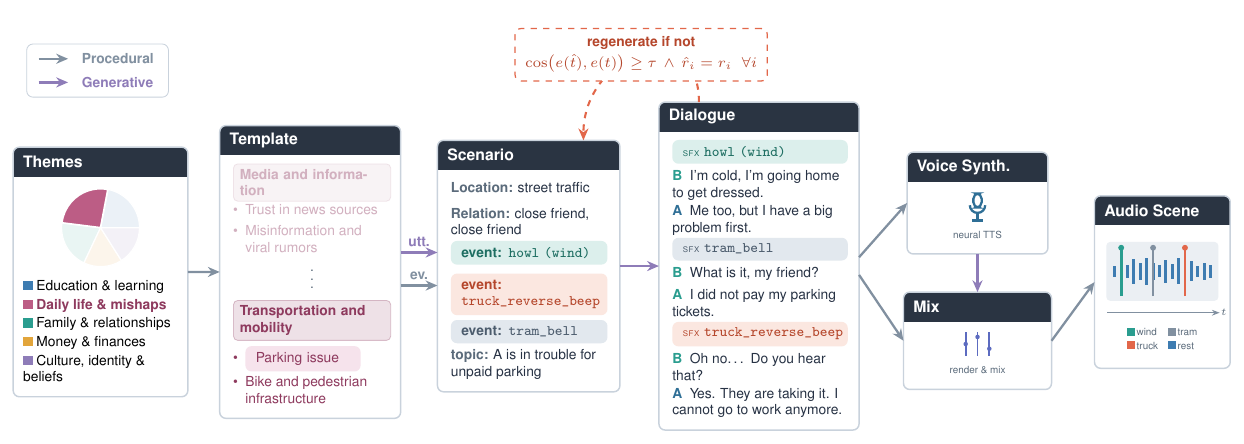}}
    \caption{The \textsc{CARES} construction pipeline: text generation 
and audio rendering. A deterministic allocator fixes the scene,
relation, speaker genders, sound events, reaction registers,
and composition. A language model fills only the conversation topic and the
dialogue. The reaction-recovery filter sits between the two phases: rejected
dialogues are regenerated until an independent evaluator recovers the intended
reactions. \emph{utt.}: utterances; \emph{ev.}: acoustic events.}
    \label{fig:pipeline}
\end{figure*}

Each scene starts from a \emph{template}, 
a short situation two adults could discuss (e.g.\ \texttt{Parking issue}) (Fig.~\ref{fig:pipeline}).
Each template is used to generate several \emph{scenarios}. 
A scenario defines: who speaks, 
a \emph{relation} (e.g. two close friends), 
the gender of each speaker, 
placed in a \emph{scene} (an everyday location such as a caf\'e or a train station),
talking about a \emph{topic}, 
during which occur audio \emph{events}, (e.g.\ a coffee machine or a door closing noise). 
Each event carries a reaction type and a register (Sec.~\ref{ssec:react}).

\subsection{Speaker reactions to acoustic events}
\label{ssec:react}
Each event is assigned one of four reaction types
(Fig.~\ref{fig:taxonomy}), defined by the participants' displayed
conduct rather than by the physical properties of the sound. They go
from a sound that takes over the conversation to one that draws no
reaction.
A \textbf{pivot} event is named and the conversation reorganizes around
it.
A \textbf{verbal} event is named in passing, and the talk resumes its
prior course within the same or the next turn.
A \textbf{behavioral} event is never named but disrupts the talk, as
when a speaker breaks off, restarts or is interrupted.
An \textbf{ambient} event draws no reaction: it is neither named nor
does it change the talk, whatever its loudness.

Each reaction type except ambient comes in a fixed set of
\emph{registers}, i.e.\ ways to react: a verbal event may be named in
passing or commented on for its loudness, a pivot may start an argument
about whether it should happen here, a behavioral event may make the speaker lose their place. We keep a
register only if the reaction would make no sense without the sound.

\textbf{Why pre-allocation matters.} If the language model chose the
content of each scenario itself, it would concentrate its choices on a
few familiar modes~\cite{zhang2025verbalized}. Our allocator instead
enforces a uniform distribution over scenes, relations, genders,
registers, per-scene event banks and compositions (how many events of
each type a scenario carries), and shuffles the order of events so that
it does not encode their type.

\subsection{Dialogue generation}
\label{ssec:dialogue}
Each scenario is turned into a textual dialogue generated as a
timeline: an ordered list of utterances (speaker A or B) and
acoustic events. The prompt enforces the reaction design: every
reacted-to sound must draw a reaction, two such sounds must draw
distinct and separately placed ones, ambient sounds must draw none. A
behavioral break must fall in the first sentence after the sound, and
the speaker never explains it, since a stated cause would replace the
sound. Other constraints are stylistic (spoken English, entity
grounding, optional inline tags such as \texttt{[laughs]}, each rendered
by the synthesizer as a vocalization, a pause or an emotion, so that no
tag is left without a trace in the audio). Dialogues run up to 13
utterances. A script verifies the rules that can
be checked automatically.
The full prompts are released with the
code.\footnote{\url{https://github.com/marcelgibier/CARES}}

\subsection{Audio rendering}
\label{ssec:audr}
Audio rendering is procedural and seeded, apart from the neural TTS voice track. The mix combines three layers:
the voice track, one sample per acoustic event, and a per-scene
ambience, low-pass filtered so it does not mask the voice. Since the
turn timings returned by the synthesizer do not match the rendered
audio, we recover word and turn boundaries with a forced
aligner~\cite{pratap2023mms}. Every boundary then receives a short gap
drawn from the same range, event or not, and each event reserves a clear
window of 0.35\,s on its loudest moment. Events are summed onto the
voice track, so the voice is only shifted, never cut, and the length of
a gap says nothing about the sound in it. A behavioral sound ends just
before speech resumes and ducks the reacting voice by up to 4\,dB, so it
explains the interruption instead of covering it. Room-impulse-response
reverberation places each scene in a plausible space, and every event
sits at least 6\,dB above the ambience bed: an annotated sound is always
audible, while levels never encode the reaction.

\section{Implementation Details}
\label{sec:implem}

\textbf{Scale.} The corpus consists of 1{,}000 templates and 10{,}000 scenarios. The scenarios cover 24 pairs
of adult roles and 13 locations (about 770 scenarios each), with 219
different sounds.

\textbf{Splits.} We define a  80\%
10\% 10\% split of the 1{,}000 templates, for the train, validation and
test sets, respectively. The split is made on templates, not on scenes and all the scenarios of a template follow it (no situation is
therefore shared across splits). The test split contains 1000 scenes, and
every number below is measured on it. The train and validation splits are released for fine-tuning and probing, the benchmark itself is zero-shot.

\textbf{Compositions.} The sound layer of a scenario is the tuple of
event counts per reaction type, $(n_{\mathrm{piv}}, n_{\mathrm{ver}},
n_{\mathrm{beh}}, n_{\mathrm{amb}})$. A scenario carries 0 to 3 events
with at most one pivot over and its
composition is drawn uniformly.

\textbf{Models.} Conversation topics come from
Qwen3-235B-Instruct~\cite{qwenteam}; dialogues, with their event
placement, from Claude Opus 5~\cite{anthropicclaude}, a different model
family from both the scenario step and the evaluator (Llama-4-Maverick-17B-128E-Instruct~\cite{llama4}). Speech is synthesized with
ElevenLabs~v3~\cite{elevenlabsv3}, which renders expressive
multi-speaker audio and interprets the inline tags.

\textbf{Audio assets.} Event takes and ambiences come from Freesound (30
clean, isolated takes per event identifier, 30 ambiences per location),
drawn at random at mix time. An ambience is trimmed or looped with
cross-fades to the length of the scene. Takes, ambiences and room
impulse responses follow the same three-way partition: 5 takes per
identifier and 5 ambiences per location are reserved for test, so no
recording heard in test occurs in train or validation.

\textbf{Mixing and export.} Loudness follows EBU R128 with fixed
targets: voice-level events at $-20$\,LUFS, set-back events at $-23$
(the layer drawn at random), scene ambience at $-32$ (jittered per
scene), and a $-23$\,LUFS master. Audio is exported as 44.1\,kHz mono
16-bit WAV with a peak limiter at $-1$\,dB.
The released set contains 10{,}000 scenes, $120.4 \pm
18.8$\,s long on average, with scenes, relations, events and
compositions uniformly distributed by construction.

\section{Validation}
\label{sec:validation}
The allocator fixes the reaction type of each sound, but the language
model may not write it as intended: a pivot can end up as a passing
mention, or a behavioral break can be explained by the speaker. The goal
of this validation is to check that each dialogue carries the reactions
it was asked for, so that the type of each sound can be recovered from
the dialogue alone, and that the final mix keeps this
dialogue intelligible.

\textbf{Reaction-recovery filter.} We enforce a per-item
property at construction time. Let $X$ be a scenario: a topic $t$ and,
for each environmental sound $i$, a gold reaction type $r_i$. An
independent evaluator $g$ is shown the full transcript of a generated
dialogue $Y = f(X)$, inline tags included, with every event marked as
\texttt{[SOUND: <id>]}. All sounds are thus visible: $g$ does not detect
them but classifies them, returning $\hat{t}$ and a type $\hat{r}_i$ per
sound. We accept $Y$ when
\begin{equation}
\label{eq:accept}
\cos\!\big(e(\hat{t}),\, e(t)\big) \ge \tau \quad\wedge\quad
\hat{r}_i = r_i \;\; \forall i,
\end{equation}
with $e(\cdot)$ a BGE-large-en-v1.5 embedding~\cite{xiao2023bge} and
$\tau$ a threshold. Sounds are matched exactly on their identifier, so
$\hat{r}_i$ is a four-way label. The criterion is strict: neither
omission nor confusion is tolerated, and failing dialogues are
regenerated. Two properties make the filter meaningful. First, $f$
(Claude Opus 5) and $g$ (Llama-4 Maverick) come from different model
families, so acceptance cannot reduces the risk of shared bias. Second, the filter
is a rejection-sampling loop: candidates are drawn until one passes, so
it selects which dialogue is kept for a given $X$ without altering the
target distribution $p(X)$.

\textbf{Topic threshold.} In Eq.~\eqref{eq:accept}, $\tau$ is the cosine
cutoff above which gold and recovered topics count as matching. BGE scores are uncalibrated, so $\tau$ is set on our data: 0.82, at the max-Youden point (ROC AUC 0.861) between gold topics paired with a loose reformulation and with another scenario's topic from the same template.

\textbf{Human validation.} Three native English speakers with a high-school education, recruited
on Fiverr\footnote{\url{https://www.fiverr.com/}}, independently labelled the sounds of 200 test scenes. They were
given the definitions of the four categories, read the reference
transcript with each event marked in place, and had no access to the
dataset label.
They agree with the released labels at $\kappa = 0.86$ and with each
other at $\kappa = 0.81$. The four categories are thus recoverable from the transcript by a
person, not only by the model that assigned them.

\begin{table*}[t]
\centering
\footnotesize
\setlength{\tabcolsep}{3pt}
\resizebox{\textwidth}{!}{%
\begin{tabular}{@{\extracolsep{\fill}}lccccccccccc@{}}
\toprule
& \multicolumn{5}{c}{\textbf{Reaction Classification}} & \textbf{Acoustic Scene} & \textbf{Audio} & \multicolumn{4}{c}{\textbf{Summary}} \\
\cmidrule(lr){2-6}
& Balanced & \multicolumn{4}{c}{Precision(\%)\,/\,Recall(\%)} & \textbf{Classification} & \textbf{Tagging} & \multicolumn{4}{c}{Recall (\%)$\uparrow$} \\
\cmidrule(lr){3-6}\cmidrule(lr){7-7}\cmidrule(lr){8-8}\cmidrule(lr){9-12}
Model & Accuracy(\%)$\uparrow$ & \textit{Pivot} & \textit{Verbal} & \textit{Behav.} & \textit{Ambient} & Accuracy (\%)$\uparrow$ & Accuracy (\%)$\uparrow$ & \textit{Pivot} & \textit{Verbal} & \textit{Behav.} & \textit{Ambient} \\
\midrule
Audio Flamingo Next & $29.7_{\pm2.0}$ & \cellcolor{red!20}50.0/0.3 & 30.6/65.0 & \cellcolor{red!20}50.0/0.7 & 41.1/52.7 & $62.5_{\pm3}$ & \cellcolor{blue!20}$34.8_{\pm2}$ & $9.5_{\pm3.7}$ & $3.3_{\pm1.7}$ & $0.5_{\pm1.0}$ & $2.5_{\pm1.5}$ \\
Qwen3-Omni          & \cellcolor{blue!20} $\mathbf{39.0}_{\pm2.0}$ & 27.0/89.5 & 33.6/41.7 & 41.5/24.2 & \cellcolor{red!20}75.0/0.5 & \cellcolor{blue!20}$\mathbf{89,6}_{\pm2}$ & \cellcolor{blue!20}$\mathbf{57.7}_{\pm2}$ & \cellcolor{blue!20}$\mathbf{22.2}_{\pm4.9}$ & \cellcolor{blue!20}$\mathbf{6.0}_{\pm2.2}$ & $\mathbf{1.6}_{\pm1.4}$ & \cellcolor{blue!20}$\mathbf{3.3}_{\pm1.7}$ \\
Kimi-Audio          & $25.6_{\pm2.0}$ & \cellcolor{red!20}27.8/1.6 & \cellcolor{red!20}29.6/1.3 & \cellcolor{red!20}22.2/5.4 & 31.2/94.1 & \cellcolor{blue!20}$78.1_{\pm2}$ & \cellcolor{blue!20}$48.6_{\pm2}$ & $6.7_{\pm3.3}$ & $1.7_{\pm1.4}$ & $0.5_{\pm1.0}$ & $0.8_{\pm1.0}$ \\
MiMo-Audio          & $32.8_{\pm2.0}$ & 40.6/40.0 & 30.4/91.2 & \cellcolor{red!20}---/0.0 & \cellcolor{red!20}---/0.0 & \cellcolor{blue!20}$81.7_{\pm2}$ & \cellcolor{blue!20}$46.5_{\pm2}$ & $12.0_{\pm4.0}$ & \cellcolor{blue!20}$4.0_{\pm2.0}$ & $1.0_{\pm1.0}$ & $2.0_{\pm1.2}$ \\
MOSS-Audio          & $31.4_{\pm2.0}$ & 17.4/99.4 & 43.7/17.9 & \cellcolor{red!20}76.5/7.0 & \cellcolor{red!20}50.0/1.4 & \cellcolor{blue!20}$71.1_{\pm3}$ & \cellcolor{blue!20}$48.5_{\pm2}$ & $5.5_{\pm3.0}$ & $1.5_{\pm1.2}$ & $0.2_{\pm0.5}$ & $0.5_{\pm0.8}$ \\
MiDashengLM         & $24.1_{\pm2.0}$ & 14.6/77.1 & 25.6/19.2 & \cellcolor{red!20}---/0.0 & \cellcolor{red!20}---/0.0 & $48.1_{\pm3}$ & $33.3_{\pm2}$ & $3.0_{\pm2.0}$ & $1.0_{\pm1.0}$ & $0.0_{\pm0.0}$ & $0.2_{\pm0.5}$ \\
\midrule
Whisper + LLM       & $36.5_{\pm2.0}$ & 59.5/23.8 & 34.6/35.4 & 36.2/22.9 & 40.3/64.0 & $64.6_{\pm3}$ & $31_{\pm2}$ & $16.5_{\pm4.1}$ & $1.5_{\pm1.0}$ & $0.7_{\pm0.7}$ & $1.1_{\pm0.8}$ \\
\bottomrule
\end{tabular}}
\vspace{-2pt}
\caption{Benchmark on the 1{,}000 test scenes. Whisper\,+\,LLM is the
text-only control. All scores in \%, chance is 25 for all accuracies.
\textbf{Balanced} accuracy is the mean of the four per-class accuracies. Recall and precision are provided for each reaction class. Acoustic scene classification and audio tagging are four-way multiple choice.
Summary recall refers to the proportion of annotated sounds of this type that are mentioned in the 200-words summary generated by the model.
($\pm$) give the 95\,\% confidence interval. Best per column in
\textbf{bold}, except for per-class precision/recall. Results significantly higher than
the text-only control are in \colorbox{blue!20}{blue}. Classes with a \textit{Reaction Classification} recall$\leq 7\%$ are shown in \colorbox{red!20}{red}.}
\label{tab:bench}
\end{table*}

\textbf{Speech intelligibility.} Events and ambiences are mixed on top of
the speech, so they could mask the dialogue and hide the reactions it
carries. To check that they do not, we transcribe the final mix with
Whisper-large-v3~\cite{radford2023whisper} and compare it with the
synthesis scripts. The word error rate is 3.3\% (95\% CI [2.8, 3.9]),
close to the 2.7\% Whisper reaches on clean read speech (LibriSpeech
test-clean): the added sounds leave the dialogue intelligible.

\section{Benchmark}
\label{sec:benchmark}

Audio--language models can in principle relate what is said to what is
heard, but no dataset could so far test whether
they relate a sound to how the speakers react to it. \textsc{CARES}
makes this test possible, since the reaction to each sound is known
before the audio exists. 

\textbf{Evaluated tasks.} The benchmark tests whether a model can tell how
speakers react to a sound, and whether hearing the audio helps beyond
the words. It has four tasks, posed as text questions on the 1000 scenes
of the test split, whose templates and recordings never appear in
training or validation. In \emph{reaction classification} (RC), the
model is told when a sound occurs and picks one of the four reaction
types, whose definitions are given in the prompt. \emph{Acoustic scene
classification} (ASC) and \emph{audio tagging} (AT) are four-way
multiple-choice questions on the location and on the sounds present,
with AT distractors taken from the same location. 

\textbf{Evaluated models.} We restrict the benchmark to open-weight models, so that every score can be reproduced from the released audio. We evaluate six recent open audio--language models
(ALMs): Audio Flamingo Next~\cite{goel2026afnext},
Qwen3-Omni~\cite{xu2025qwen3omni}, Kimi-Audio~\cite{kimiteam2025kimiaudio},
MiMo-Audio~\cite{xiaomi2025mimoaudio}, MOSS-Audio~\cite{yang2026mossaudio}
and MiDashengLM~\cite{dinkel2025midasheng}. We compare them with a
text-only system that never hears a sound: Whisper-large-v3 transcribes
the speech and a text model (Qwen3.6-35B-A3B~\cite{qwenteam}) answers
from the transcript. If it matches the ALMs, a task is solved from the
words alone.

\section{Results}
\label{sec:results}

\textbf{Do the models recognize the sounds?} 
As shown in the columns "Accoustic Scene Classification" and " Audio Tagging" in Table \ref{tab:bench}, most models are able to classify both the nature of the audio scene (with between 48\% and 90\% accuracy over 4 classes) and, in a lesser measure, the audio events (with between 35\% and 58\% accuracy over 4 classes), topped each time by the \emph{Qwen3-Omni} model.
However, this perception is not purely based on audio processing, as the contrastive text-only experiment shows similar performance, notably for the \emph{Audio Flamingo Next} and \emph{MiDashengLM} models. 
The high accuracy on the audio scene classification from text ($65\%_{\pm3\%}$) is easily explained as the audio scenes are selected to match the topic of the conversation, while the non-random audio tagging results from text ($31\%_{\pm2\%}$) may come from some \textit{pivot} and \textit{verbal} dialogues mentionning explicitely the nature of the sound.


\textbf{Can the models tell how/if the speakers react to the sounds?}
As shown in the \textit{Reaction Classification} column of Table \ref{tab:bench}, the best audio model reaches 39.0\% of balanced accuracy, close to the 36.5\% of the text control. 
Every ALM has at least one class that gets totally ignored (Recall$\leq 7\%$), each collapsing onto a different set: MOSS-Audio answers pivot almost always, Kimi-Audio
ambient, and MiMo-Audio and MiDashengLM never answer behavioral or
ambient. 
The text control is the only system that ignores none of the classes. 


We also ran reaction classification on the same scenes with the
annotated sound muted : no ALM's balanced accuracy moved
significantly. The models rely on what is said, not on the audio event. What the models use is in the speech, not in the
sound.

\textbf{The importance of the sound in the summarization}
From the Summarization recall columns of Table \ref{tab:bench}, we see no model judged the audio events important enough to be recorded in the summary of the conversation.
Even when an audio event becomes the topic of the talk, the best model keeps 22\% of them, while the text-only control keeps more (16.5\%) than five of the six ALMs. Recall follows the same
order as above. Even when an unnamed sound broke the conversation, it
rarely reaches the summary: the ALMs keep at most 3.3\% of unnamed
sounds, against about 1\% for the control, and no more when the sound
broke the talk than when nobody noticed it. In summaries as in classification,
what counts is what the speakers say about a sound, not what it does to the conversation.

\section{Conclusion}
\label{sec:conclusion}

Automatic audio scene description lacks a ground truth for which audio
events matter. We define salience by the reaction a sound draws, fix
this ground truth first, and generate the audio around it. The
resulting corpus, \textsc{CARES}, contains 10{,}000 two-speaker scenes.
Audio--language models recognize its sounds, but tell how speakers react
no better than a text-only system, and their summaries rarely mention a
sound nobody named. The open problem is not recognizing a sound, but
linking it to what the talk does with it.

\textbf{Limitations.} The scenes are synthetic English dialogues,
checked on transcripts rather than audio. Language models write and
check the dialogues and judge the summaries, which may favour reactions
they read easily. Apart from Qwen3-Omni, we cannot tell whether the
models' collapse comes from the audio or from the task format.

\section{Acknowledgements}
This work was performed using HPC resources from GENCI IDRIS
(Grant 2025-AD011014982R1).

\bibliographystyle{IEEEbib}
\bibliography{refs}

\end{document}